\documentclass[aps,pre,floats,twocolumn,showpacs,superscriptaddress]{revtex4}
\usepackage{graphicx}  
\usepackage{dcolumn}   
\usepackage{bm}        
\usepackage{amssymb}   
\usepackage{amsmath}
\usepackage{bbm}
\usepackage{doi}
\usepackage{hyperref}
\usepackage{epsfig}
\usepackage[section]{placeins}
\usepackage{float}

\usepackage{color}
\definecolor{orange}{RGB}{255,127,0}

\newcommand{\pons}[1]{\textcolor{red}{#1}}

\usepackage{ulem}

\begin{document}

\title{Consistency of heterogeneous synchronization patterns in complex weighted networks}
\author{D. Malagarriga}
\thanks{Electronic address: \texttt{daniel.malagarriga@upc.edu}; Corresponding author}
\affiliation{Departament de F\'isica, Universitat Polit\`ecnica de Catalunya. \\ Edifici Gaia, Rambla Sant Nebridi 22, 08222 Terrassa, Spain.}
\affiliation{Center for Genomic Regulation (CRG). \\ Barcelona Biomedical Research Park (PRBB), Dr. Aiguader 88, 08003 Barcelona, Spain.}
\author{A.E.P. Villa}
\affiliation{Neuroheuristic Research Group, Faculty of Business and Economics, University of Lausanne. \\ CH-1015 Lausanne, Switzerland.}
\author{J. Garcia-Ojalvo}
\affiliation{Department of Experimental and Health Sciences, Universitat Pompeu Fabra. \\ Barcelona Biomedical Research Park (PRBB), Dr. Aiguader 88, 08003 Barcelona, Spain.}
\author{A.J. Pons}%
\affiliation{Departament de F\'isica, Universitat Polit\`ecnica de Catalunya. \\ Edifici Gaia, Rambla Sant Nebridi 22, 08222 Terrassa, Spain.}
\email{daniel.malagarriga@upc.edu}

\date{\today}
\begin{abstract}  
We show that {subsets} of interacting oscillators may synchronize in different ways within a single network. This diversity of synchronization patterns is promoted by increasing the heterogeneous distribution of coupling weights and/or asymmetries in small networks. We also analyze {\it consistency}, defined as the persistence of coexistent synchronization patterns regardless of the initial conditions. Our results show that complex weighted networks display richer consistency {than regular networks}, suggesting why certain functional network topologies are often constructed when experimental data are analyzed.
\end{abstract}
\maketitle
\section{Introduction}
Certain dynamical systems, which display oscillatory behavior in isolation, may display a wide repertoire of dynamical evolutions due to the coupling with their neighbors when embedded in networks of similar complex items~\citep{pikovsky2003synchronization}. For instance, the interaction of rhythmic elements may entail an adjustment of their oscillatory dynamics to finally end up in a state of (dynamical) {\it agreement} or {\it synchronization}~\cite{strogatz2003sync,strogatz2001,Glass2001}.
When coupling is strong, the oscillators in a network usually synchronize in a particular collective
oscillatory behavior. However, {for more moderate coupling intensity,} this relationship may
also be inhomogeneous, i.e., certain
oscillators may synchronize whereas others may not~\cite{Abrams2004,omelchenko2011loss,Pecora2014,Schmidt2014,Schöll2016}. The specific patterns of
synchronization, thus, may provide information about the underlying dynamics and {about} the
{coupling} between the dynamical elements forming {the}
network. Hence, {by analyzing all the synchronization relationships in a network,} a
better characterization of the system can be achieved{. This characterization} might be
of crucial importance when the
details of the contacts between the oscillators is not available. 

A fundamental question in network science is the relationship 
between network dynamics and network structure.
In the past, studies of the synchronization patterns in networks of oscillators 
were mainly aimed at describing the conditions 
associated with the emergence of specific synchronization patterns
in all the nodes
	\cite{Chavez2005}. 
In the particular case of complex networks of coupled nonlinear oscillators,
recent studies have provided evidence that 
it is possible to identify an appropriate interaction regime 
that allows to collect measured data
to infer the underlying network structure
based on time-series statistical similarity analysis
	\cite{Tirabassi2015}
or connectivity stability analysis
	\cite{Lin2015}.
In real-life systems,
such as
ecological networks
	\cite{blasius_complex_1999},
brain oscillations
	\cite{Robinson1998,Hill1997165,Cabessa2014,Malagarriga2015segregation}
or climate interactions
	\cite{Deza2015}, 
various types of complex synchronized dynamics have been observed.
%
%
%
Therefore, such a diversity in dynamical relationships between the nodes
endows a network with stability, 
flexibility and robustness against perturbations
	\cite{zanette2004propagation}.

We show that several types of {stable} synchronization patterns may coexist
depending on the topology and on the distribution of coupling strengths within a network.
We relate the capacity of a network 
to display the same coexistence pattern regardless of the initial conditions 
with its {\it consistency}.
Finally, we suggest that the retrieval of a network structure
from its dynamics is very reliable 
when the coexisting synchronization patterns
{are consistent.}
 
\section{Coexistence of synchronizations}

\begin{figure*}[ht!]

\centering
{\includegraphics[width=\textwidth, keepaspectratio,clip=True]{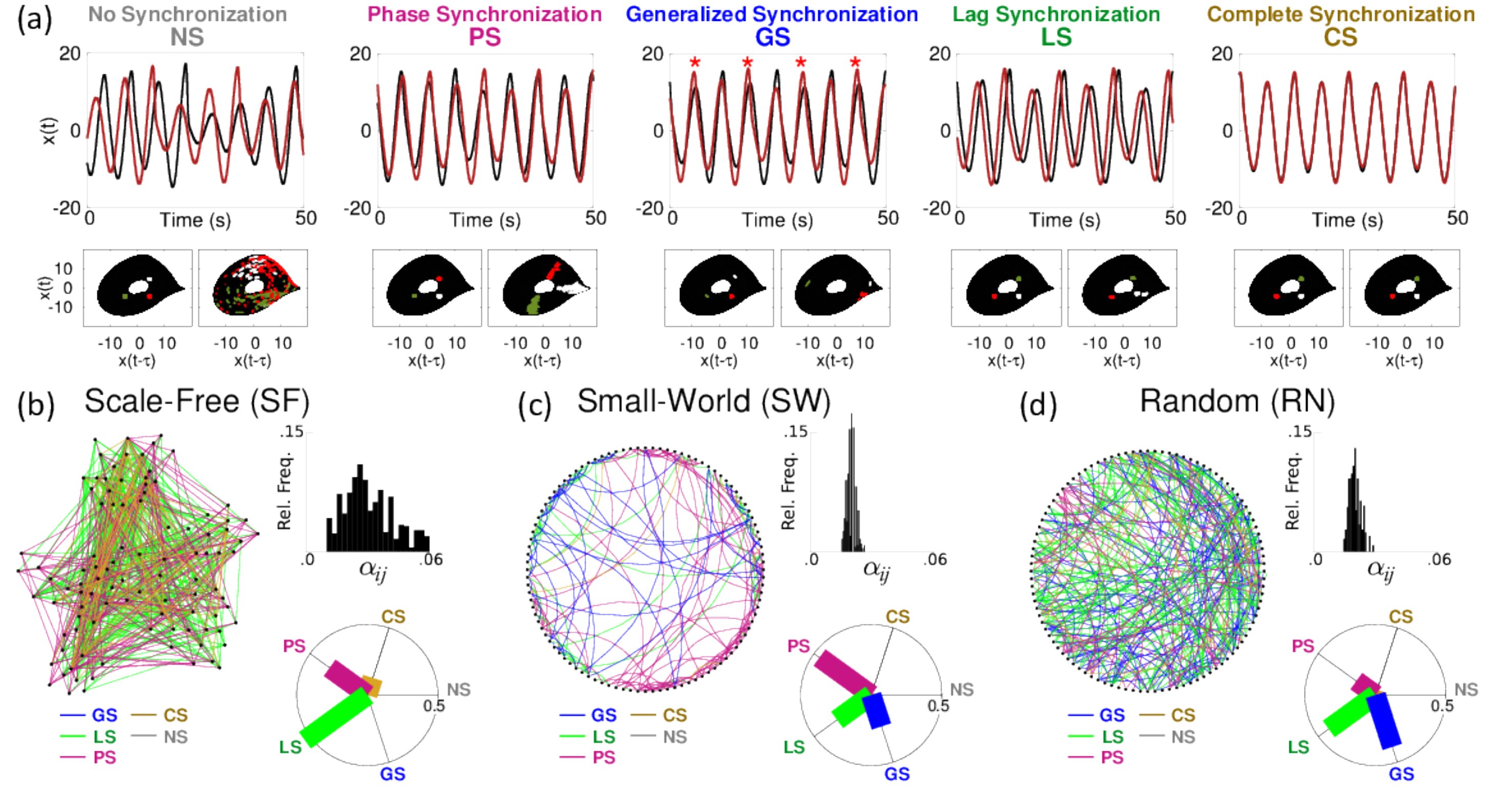}}
\caption[Heterogeneous synchronization patterns in complex weighted networks.]{(Color online)  
{\bf Heterogeneous synchronization patterns in complex weighted networks.} 
(a) Examples of synchronization patterns
(no synchronization NS, phase synchronization PS, generalized synchronization GS, lag
synchronization LS and complete synchronization CS) displayed by {bidirectionally}
coupled R\"{o}ssler
oscillators{. The upper panels show examples of $x_i(t)$ timetraces for each synchronization
pattern and the lower panels show examples of the corresponding {delay-embedding}
plot~}\citep{Moskalenko2012}. 
$\tau$ is the delay time for maximal cross-correlation in LS and PS.
Examples of (b) a scale-free (SF) network ($K=0.4$), 
(c) small-world (SW) network with low rewiring probability ($K=0.1$),
and ({D}) random (RN) ($K=0.1$) of coupled R\"{o}ssler oscillators
displaying heterogeneous synchronization patterns. All networks have $N=100$ nodes. $K$ is a global coupling parameter controlling the maximum coupling strength between two adjacent nodes (see Eq.~\ref{eq:Eq_Rossler}).
For each type of network the right panels show 
the distribution of the coupling strengths $\alpha_{ij}$ between pairs of nodes (upper panel)
and the distribution of the synchronization patterns (polar histogram, lower panel). Each link is color-coded so as to show which synchronization pattern is displayed by each pair of oscillators within the network (NS, PS, GS, LS, see left bottom panel). 
}
\label{Chap4:fig_1}

\end{figure*}

Consider two dynamical systems, ${\bf{x}}$ and ${\bf{y}}$, whose temporal evolutions are generally defined by  
${\bf\dot{x}}(t)=\bf{F}(\bf{x}(t))$, ${\bf\dot{y}}(t) = \bf{G}({\bf y}(t))$ in isolation. Assuming a bidirectional coupling scheme, the coupled system reads:

\begin{eqnarray}
{\dot{\bf x}}(t) = {\bf F}({\bf x}(t)) + \hat{\bf{C}}({\bf y}(t)-{\bf x}(t)), \nonumber \\
{\dot{\bf y}}(t) = {\bf G}({\bf y}(t)) + \hat{\bf{C}}({\bf x}(t)-{\bf y}(t)).
\end{eqnarray}
${\bf x}(t)$ and ${\bf y}(t)$ are the state vectors of the systems, ${\bf F}$ and ${\bf G}$ are
their corresponding vector fields and $\hat{{\bf C}}$ is a {$n \times n$} matrix that
provides the coupling characteristics between the sub-systems. When coupling is strong enough
{and these dynamical systems are oscillators,} the
synchronization relationships that can be established between {them} can be categorized in
four {types} (see time traces in
Fig.~\ref{Chap4:fig_1})~\citep{Uchida2005203,Xiao-Wen2007}:

\begin{itemize}
\item Phase synchronization (PS) appears if the functional relationship between the dynamics of two
oscillators preserves a bounded phase difference~\citep{Rosenblum1996}, with their amplitudes being
largely uncorrelated. This can be exemplified by the relationship $|n\phi_1-m\phi_2|<
\mathrm{const}$, with $\phi_{1,2}$ being the phases of the two coupled oscillators. 
\item Generalized synchronization (GS) is observed if a complex functional relationship is
established between the oscillators~\citep{Abarbanel1996}, e.g. {${\bf{ y}}(t) = H[{\bf x}(t)]$, where
$H[{\bf x}(t)]$} can take any form other than identity. It can be thought to be a
generalization of CS for non-identical oscillators. 
\item Lag synchronization (LS) appears when the
amplitude correlation is high while at the same time there is a time shift in the dynamics of the
oscillators~\citep{Rosenblum1997}, ${\bf{y}}(t) = {\bf{x}}(t-\tau)$, with $\tau$ being
a lag time.
\item Complete synchronization (CS) is observed when the coupled oscillators are identical or almost
identical~\citep{Boccaletti2002}, and ${\bf x}(t) = {\bf y}(t)$ for a sufficiently large coupling
strength $\hat{\bf{C}}$. 
\end{itemize}

There are several analysis techniques that can be used to assess the emergence of each of the mentioned synchronization motifs. Here we combine three of them: cross-correlation (CC), Phase-Locking Value (PLV) and the Nearest-Neighbor Method (NNM). CC computes the lagged similarity between two signals, which provides a notion of the  amplitude resemblance over time. Therefore it allows to {identify} whether CS or LS are established between two time traces. On the other hand, PLV makes use of the Hilbert transform of a signal to retrieve a phase $\phi$ and compute the time evolution of the difference in the phases of two oscillators, i.e. $\phi_1(t)-\phi_2(t)$~\citep{Lachaux1999}, as:
\begin{equation}
PLV_t = \frac{1}{N} \left | \sum_{n=1}^N \langle e^{ i\Delta\phi_{12}(t,n)}\rangle_t\right |, 
\end{equation}
where $\Delta\phi_{12}(t,n)$ is the evolution of the difference between the phases of oscillators 1
and 2, $N$ is the number of trials and $\langle...\rangle_t$ denotes temporal average. {This measure can assess, when combined with low CC,} the emergence of PS between two oscillators. Finally the NNM
takes points in the phase space of {each} oscillator and {characterizes} their relative evolution~\citep{Boccaletti2002}. {This method} allows to visualize and exemplify each
synchronization motif (see examples in Fig.~\ref{Chap4:fig_1}(a), lower panels). With this set of
analysis techniques, here we study the dynamics of 
networks of coupled R\"{o}ssler oscillators~\citep{ROSSLER1976397} arranged in complex weighted topologies 
--~random (RN)
	\citep{ErdosRenyi1959},
small-world (SW)
	\citep{watts1998cds} and 
scale-free (SF)
	\citep{Barabasi15101999}~--.

Firstly, we set the dynamics of each node $i$ to follow the R\"{o}ssler equations:
\begin{align}
\begin{split}
\dot{x}_{i} &= -\omega_{i}y_{i} - z_{i} +K\sum_{j=1,j\neq i}^{N_{neigh}} \alpha_{ij}(x_j-x_i), \\
\dot{y}_{i} &= \omega_{i}x_{i} + ay_{i}, \\
\dot{z}_{i} &= p + z_{i}(x_{i}-c),
\end{split}
\label{eq:Eq_Rossler}
\end{align}
where $K$ is a global parameter controlling the maximum coupling strength between two nodes and $\omega_i$ is the natural frequency of the node $i$, 
which is normally distributed with average $\langle\omega\rangle = 1$ 
and standard deviation $\sigma_{\omega} = 0.02$. An isolated node with R\"{o}ssler dynamics 
can display periodic, quasi-periodic or chaotic dynamics,
and we choose $a = 0.15$, $p = 0.2$ and $c = 10$ 
to set the oscillators into a chaotic regime 
	\cite{Boccaletti2002}. 
We set the coupling weights to depend on the number of neighbors of each node, if not specified otherwise, as:
\begin{equation}
\alpha_{ij} = \frac{1}{\sqrt{\textrm{deg}(v_i)\textrm{deg}(v_j)}},
\end{equation}
for $i\neq j$, where deg$(v_i)$, deg$(v_j)$ are the degrees (number of coupled neighbors) of two coupled nodes $v_i$, $v_j$. We study regular and complex topologies of progressively larger networks.

{Figure{~\ref{Chap4:fig_1}(b-d)} shows} the distribution of synchronizations in three
prototypical networks (composed of $N=100$ nodes), namely, SF, SW and RN, alongside with their
weight distributions (relative frequency of {$\alpha_{ij}$}) and the distribution
of synchronizations within each network. All three networks are located in a region of the coupling
parameter space which allows a complex distribution of synchronizations. {In this sense, the SF network shows clusters of PS, LS and CS, and SW and RN
networks show clusters of PS, GS and LS.} However, such distribution is very sensitive to the
coupling characteristics and the underlying topology. Therefore, we want to {understand better} which are the conditions for the non trivial distribution of
synchronizations to appear by analyzing the interaction of the nodes' dynamics and the topological
properties of the networks in which they are embedded. Hence, {the following} question
arises: \textit{What are the conditions that are suitable for {different} synchronization patterns
{to coexist within}
a network?}

\section{Consistency of synchronizations}

The heterogeneous synchronization motifs that emerge in complex networks are an excellent probe to detect functional connectivity between the oscillators in a network. Besides, if these motifs are dynamically stable, we can identify synchronized states that show up recurrently even when initial conditions change, thus becoming an {\it invariant} feature of the dynamics of the network. In this section we study which are the conditions for which the same synchronization patterns persist in time for varying initial conditions.

\begin{figure}[ht!]

\centering
{\includegraphics[width=0.9\columnwidth, trim=0cm 0cm 0cm 0cm, keepaspectratio,clip=True]
{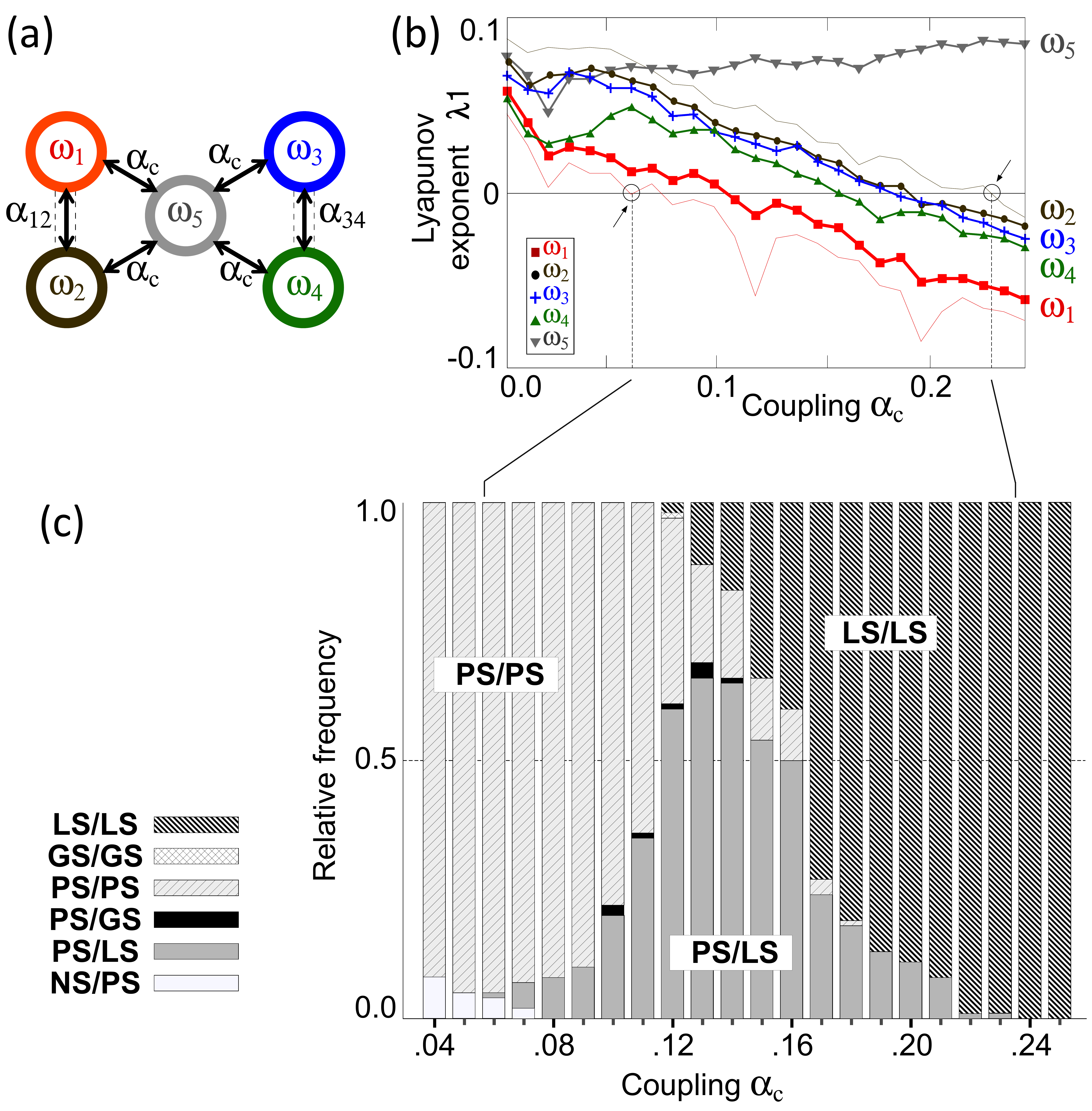}}
\caption[Dependence of the coexistence of synchronization patterns on the Lyapunov Exponents.]{(Color online)   
{\bf Dependence of the coexistence of synchronization patterns on the Lyapunov Exponents.} 
(a) Simple weighted network formed by two pairs of peripheral nodes connected to a central node.
The couplings between peripheral node pairs are $\alpha_{12}=0.05$ and $\alpha_{34}=0.03$. 
(b) For each node dynamics the curves show the mean value (computed over 100 runs with random initial conditions) 
of the maximum Lyapunov exponent ($\lambda 1$) 
as a function of the strength of coupling $\alpha_c$ of all nodes with the central node. 
The lowest thin curve corresponds to the lowest values of $\lambda 1$ for node $\omega_1$
computed independently for each value of $\alpha_c$.
This curve crosses the zero line at  $\alpha_c = 0.06$, as indicated by an arrow and a vertical dotted line. 
The uppermost thin curve corresponds to the largest values of $\lambda 1$ for node $\omega_2$.
This curve crosses the zero line at  $\alpha_c= 0.23$, as indicated by an arrow and a vertical dotted line. 
(c) Histogram of the occurrences of the synchronized patterns for each peripheral node pair in the network (1-2 and 3-4).
Notice that in the interval $\alpha_c \in [0.06,0.23]$ 
several synchronization patterns may coexist for the same coupling $\alpha_c$,
depending only on the randomly chosen initial conditions.
}
\label{Chap4:fig_2}

\end{figure}

Our first example of coexistence of synchronizations is studied in a very simple weighted network formed by two pairs of nodes connected
bidirectionally with a fifth node (see Fig.{~\ref{Chap4:fig_2}(a)}, Eqs.~({\ref{eq:Eq_Rossler}})). The oscillators only differ on the frequencies, $\omega_i$, which are the following: $\omega_1=0.930$, $\omega_2=0.967$, $\omega_3=0.990$, $\omega_4=0.950$, $\omega_5=0.970$. 
After fixing two different initial synchronized states for the two couples of peripheral nodes we change the synchronization coexistence within the network, and its stability, by increasing the bidirectional coupling $\alpha_c$ with the central node. Notice that we change the synchronization states without changing $\alpha_{1,2}$ and $\alpha_{3,4}$ (the peripheral nodes' coupling strengths). 

Since we deal with non-identical oscillators, there is no global synchronization manifold and, therefore, an {\it analytical} stability analysis of the whole system cannot be performed. However, the evolution of the coexistence of synchronized states in terms of $\alpha_c$ may be tracked numerically by considering in detail the values of the {\it conditional} Lyapunov Exponents (LEs, $\lambda 1_{\omega_{i}}$)\citep{Boccaletti2002,Moskalenko2012} when $\alpha_c$ changes (see Fig.{{~\ref{Chap4:fig_2}(b)}}), which might indicate the onset of a different synchronization motif.
Lyapunov exponents are a qualitative measure that characterize the stability and instability of the evolution of a dynamical system with respect to varying initial conditions. Briefly, if we assume that a dynamical system is described as $\dot{{\bf x}} = f({\bf x})$ with $t>0$ and initial conditions ${\bf x}(0) = {\bf x}_0 \in \mathbb{R}^n$, we can derive the variational equation of the system:
\begin{equation}
\dot{Y} = J({\bf x}(t))Y, \quad Y(0)=I_n,
\end{equation}
where $I_n$ is a {$n \times n$} identity matrix and $J({\bf x})=\partial f({\bf
x})/\partial {\bf x}$ is the Jacobian matrix of $f$. We then consider the evolution of an
infinitesimal parallelepiped in the phase space $[p_1(t),...,p_n(t)]$, with axis $p_i(t)=Y(t)p_i(0)
\:\textrm{for}\:i=1,...,n$, where $[p_1(0),...,p_n(0)]$ is an orthogonal basis of $\mathbb{R}^n$.
The long-time sensitivity of the flow ${\bf x}(t)$ with respect to initial conditions $\bf{x}_0$ at
the directions $p_i(t)$ is determined by the expansion rate of the length of the $i$th axis with
respect to the orthogonal basis $p_i(0)$, and is given by:
\begin{equation}
\lambda_i = \lim_{t \to \infty}	\frac{1}{t} ln \frac{\vert\vert{p_i(t)}\vert\vert}{\vert\vert{p_i(0)}\vert\vert} \quad i=1,...,n,
\end{equation}
which corresponds to the Lyapunov spectrum $\{\lambda_i\}$.

For the coupled case, let us first suppose that we have two oscillators, ${\bf x}(t)$ and ${\bf u}(t)$ of dimensions $N_x$ and $N_u$, respectively. For an unidirectional coupling scheme, in which ${\bf x}(t)$ drives ${\bf u}(t)$, we can consider the presence of a time-dependent functional relationship
\begin{equation}
{\bf u}(t) = {\bf F}[{\bf x}(t)].
\label{Lyapunov}
\end{equation}
The dynamics of this coupled drive-response system is characterized by the Lyapunov exponent spectra $\lambda^x_1\geqslant\lambda^x_2\geqslant...\geqslant\lambda^x_{N_x}$ and $\lambda^u_1\geqslant\lambda^u_2\geqslant...\geqslant\lambda^u_{N_u}$, with the {latter} being conditional Lyapunov exponents. In this sense, the rate of convergence or divergence of the trajectory of oscillator ${\bf u}$ towards the trajectory defined by oscillator ${\bf x}$ is given by $\lambda^u_1$: if $\lambda^u_1>0$ the trajectories diverge, whereas if $\lambda^u_1<0$ they converge.

Since throughout this manuscript we consider a mutual coupling scheme, Eq.~(\ref{Lyapunov}) no longer holds for all time $t$, but rather its implicit form ${\bf F}[{\bf x}(t),{\bf u}(t)]=0$. However, locally (i.e. for $t^*-\delta < t < t^*+\delta$, with $\delta$ being infinitely small), the implicit function theorem~\citep{Jittorntrum1978} allows to write ${\bf x}(t^*) = \hat{{\bf F}}[{\bf u}(t^*)]$ or ${\bf u}(t^{**}) = \tilde{{\bf F}}[{\bf x}(t^{**})]$, for other moments in time $t$. Therefore, without loss of generality, the spectrum of Lyapunov exponents can be computed in terms of the trajectory defined by one of the mutually coupled oscillators, either ${\bf u}$ or ${\bf x}$, as in the unidirectional coupling case. In what follows we consider the evolution of the flow of the trajectories of the coupled R\"ossler oscillators with respect to the trajectory defined by one of the oscillators in the networks. This calculation allows to estimate whether such trajectory is attractive (i.e. neighboring 
oscillators converge to it and therefore synchronize) or repulsive  (i.e. neighboring oscillators diverge from it and desynchronize in amplitude).

Figure{~\ref{Chap4:fig_2}(b)} shows that, in terms of $\alpha_c$, three different regions may be defined for the 5 (realization-averaged) largest LEs $\lambda1_{\omega_i}$:

\begin{itemize}
\item In the first region ($0<\alpha_c<0.06$) all the largest LEs are positive. The pairs 1-2 and 3-4 are mostly in PS. When increasing $\alpha_c$ in this region, peripheral nodes become PS with the central node until the first 0 crossing of $\lambda1_{\omega_i}$ (light red line), which defines the onset for GS for pair 1-2 (vertical dashed line, first arrow, $\alpha_c=0.07$). 
\item The second region ($0.07<\alpha_c<0.23$, in between dashed lines) sets a cascade of coexistence of synchronization regimes, i.e. successive zero-crossings of LEs determine the onset of GS and LS between the nodes: $\alpha_c=0.14$ defines the onset of LS between oscillators 1-2 while 3-4 remain PS.  $\alpha_c=0.16$ marks the subsequent GS onset between oscillators 3-4 while maintaining oscillators 1-2 in LS. Notice that the heterogenous pattern PS/LS is the most frequently observed.
Pattern PS/GS was rare and pattern GS/LS was never observed.
\item In the third region, after $\alpha_c=0.23$, there is the onset of LS for the whole network.
\end{itemize}

Figure{~\ref{Chap4:fig_2}(c)} shows the histogram of the occurrence of each pair of synchronized states between nodes 1 and 2 or 3 and 4, computed using CC, PLV and NNM: in the coexisting region, there exist extended $\alpha_c$ values for which pairs 1-2 and 3-4 are, simultaneously, in two different synchronized regimes, e.g. 1-2 are in LS meanwhile nodes 3-4 are in PS. Therefore, two synchronized states can coexist in the network.

\begin{figure}[ht!]

\centering
{\includegraphics[width=0.9\columnwidth, trim=0cm 0cm 0cm 0cm, keepaspectratio,clip=True]
{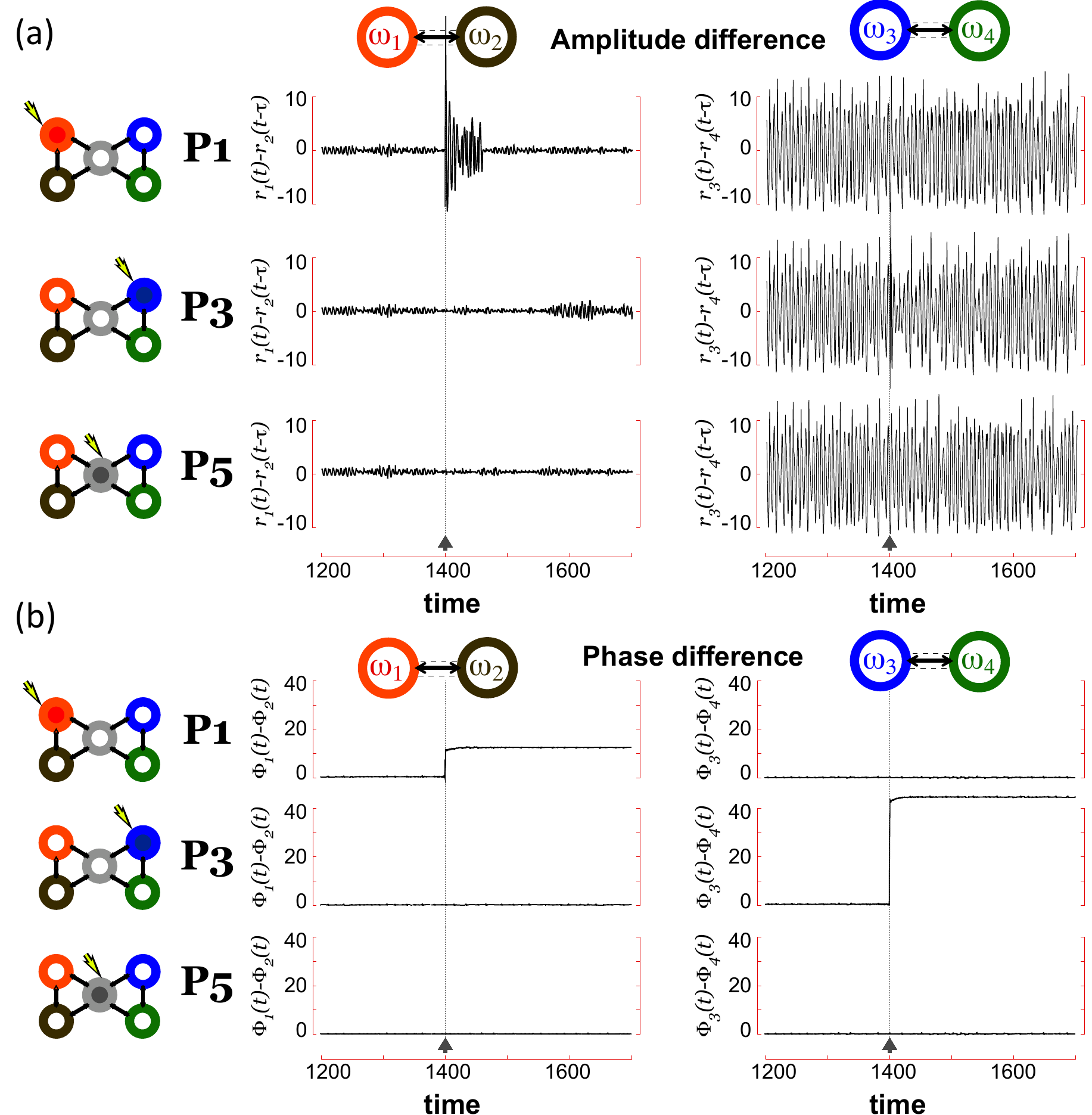}}

\caption[Robustness against perturbations.]{(Color online) 
{\bf Robustness against perturbations.} (a) We perform perturbations on {nodes} $\omega_1$ (P1),  $\omega_3$ (P3) and $\omega_5$ (P5) so as to track the amplitude difference of the two pairs of peripheral nodes. In this regard the amplitude difference is kept bounded for the two peripheral nodes, which are in LS (pair 1-2) and PS (pair 3-4), respectively ($\alpha_{1,2}=0.05$, $\alpha_{3,4}=0.03$, $\alpha_c=0.13$). Notice the divergence in amplitude and phase evolution depending on the perturbation site, indicating that the system is globally connected and senses small perturbations. (b) The phase difference suffers a sudden increase but is also kept constant in time when perturbing the same nodes as in caption (a).
}
\label{Chap4:fig_3}

\end{figure}

The relationship between the LEs exponents of this small network system and the statistical occurrence of synchronizations allows to show that the synchronization patterns obtained are stable and robust against perturbations. In order to qualitatively prove it, we have perturbed different nodes and tracked the evolution of the amplitude and phase differences between adjacent neighbors. Figure~\ref{Chap4:fig_3}(a-b) shows these time evolutions. In the coexistence regime, when the PS/LS situation is dominant ($\alpha_c=0.13$), the stable synchronized dynamics forces the pair 1-2 (in LS) to return to its bounded amplitude difference when perturbing node $\omega_1$, whereas phase differences suffer an abrupt change but do not increase for both pairs 1-2 and 3-4. Besides, perturbations in all sites of the network are sensed by all nodes, as it can be seen from the distinctive post-perturbation time evolutions of both amplitude and phase differences. Therefore, the nodes within the network do not evolve as 
isolated entities.\\

The cascade of zero-crossings of the LEs in terms of $\alpha_c$ can be expanded or squeezed by increasing or decreasing the symmetries of the system, and therefore the range of $\alpha_c$ values for which coexistence appears. For a completely symmetrical system, i.e., equal governing equations for all the nodes in a symmetric network, there are abrupt transitions to synchrony~\citep{Leyva2012}, without coexistence. Symmetry can be broken in a controlled way by means of a parameter governing the dynamics (e.g. oscillatory frequency), a parameter responsible for the topological characteristics of the network (e.g., clustering) or both features. In such scenarios different motifs of synchronized dynamics may show up, but they are restricted to a tiny region of the parameter space and, thus, appear to be spurious. We break the symmetry by adding mismatches between the frequencies of the oscillators and by increasing the heterogeneity of the nodes' degrees as well as the coupling values $\alpha_{ij}$. We show that 
symmetry crucially determines the extent of the coexistence region in the previous small network and its associated consistency. 

Figure{~\ref{Chap4:fig_4}(a) shows the  motif studied previously, but with different coupling strengths between peripheral nodes; $\alpha_{1,2}$ is now one order of magnitude smaller than $\alpha_{3,4}$ (see caption of Fig.~\ref{Chap4:fig_4}), making this motif more asymmetrical in terms of coupling strength. Again, we have tracked the evolution of the LEs in this case for increasing $\alpha_c$ values. 

Firstly, for $\alpha_c=0$, nodes 1-2 are in PS meanwhile nodes 3-4 are in GS -- i.e. a coexistence situation --. As can be seen in Fig.{~\ref{Chap4:fig_4}(a)}, for different initial conditions zero-crossings of LEs appear along an extended $\alpha_c$ value region. In this case, the coexistence region for peripheral nodes 1-2 and 3-4 spans from $\alpha_c=0$ to $\alpha_c=0.20$. The third panel in Fig.{~\ref{Chap4:fig_4}(a)} shows a plot of the relative frequency of synchronizations found for each pair of nodes in the small motif for all the realizations of the initial conditions performed and for $\alpha_c=0.04$ (three pointed star) or $\alpha_c=0.18$ (dotted circle). Consistently, each pair of peripheral nodes lies in the same synchronization state for any of the imposed initial conditions for the values $\alpha_c=0.04$ and $\alpha_c=0.18$, whereas only for $\alpha_c=0.04$ the rest of pairs display a single synchronization pattern (NS proportion not shown). Consequently, we define {\it consistency} as the capacity of a network to display the same coexistence pattern regardless of 
the initial conditions, as with $\alpha_c=0.04$.

{Figure{~\ref{Chap4:fig_4}(b)} shows a more
symmetric} network, in terms of coupling strength $\alpha_c$. Such relay
configuration is less prone to synchronize for small coupling strengths and, therefore, larger
$\alpha_c$ values are required to set synchronized states (see inset $\alpha_c=0.18$). However, the
coexistence region is also narrow and the LEs randomly cross the 0 value, which implies that the
consistency will be low. This is shown in Fig.{~\ref{Chap4:fig_4}(b)} lower right panel, in which
the relative frequency {of synchronizations} plot shows no large predominance of a single
synchronization motif{ for a given pair of nodes. Therefore, the synchronization pattern does not
correlate with the different pairs of nodes}. 

Figure{~\ref{Chap4:fig_4}(c)} shows an all-to-all small network in which all edges are weighted by the control parameter $\alpha_c$. In this case the network topology and the coupling strength distribution make this network is more symmetrical. Accordingly, the $\alpha_c$ range for which coexistence exists is narrower with respect to the previous studied motifs. This reduction of the area of coexistence has implications in the consistency of synchronizations: {zero-crossings} of LEs are randomly distributed in a tiny range of $\alpha_c$ and, so, coupled pairs in the network do not consistently lay in the same synchronized state for different initial conditions (see Fig.{~\ref{Chap4:fig_4}(c)} right panel).

Overall, by gathering the results of the coexistence and the consistency phenomena, we show that network symmetries govern the synchronization dynamics emerging from a system of coupled dynamical units~\citep{Chavez2012}. In this regard, clusters of synchronizations dynamically emerge thanks to symmetry breaking (with respect to the topology, the system parameter values or both) and the statistics of the synchronization dynamics strongly depend on the type of symmetry breaking. 

We now define synchronization clusters considering characteristic functional relationships between
the coupled elements. Thus, by labelling each of the functional relationships (NS, PS, GS, LS, CS)
{between nodes} one can have a better characterization of the global behavior of the
system {than considering only one of the relationships}. Added to this, we have shown that
there is a dependence of the statistics of coexistence on the underlying network. Therefore, when
extracting {\it functional networks} from the statistics of synchronization, we will get the most
{\it consistent} structural sub-network. {L}ess consistent sub-networks, even though they
can be coupled, show up as disconnected functionally. We will use this feature to infer
the characteristics of structural networks from the constructed functional networks.
\section{Reconstruction of consistent networks}
\begin{figure}[ht!]

\centering
{\includegraphics[width=\columnwidth, trim=0cm 0cm 0cm 0cm, keepaspectratio,clip=True]
{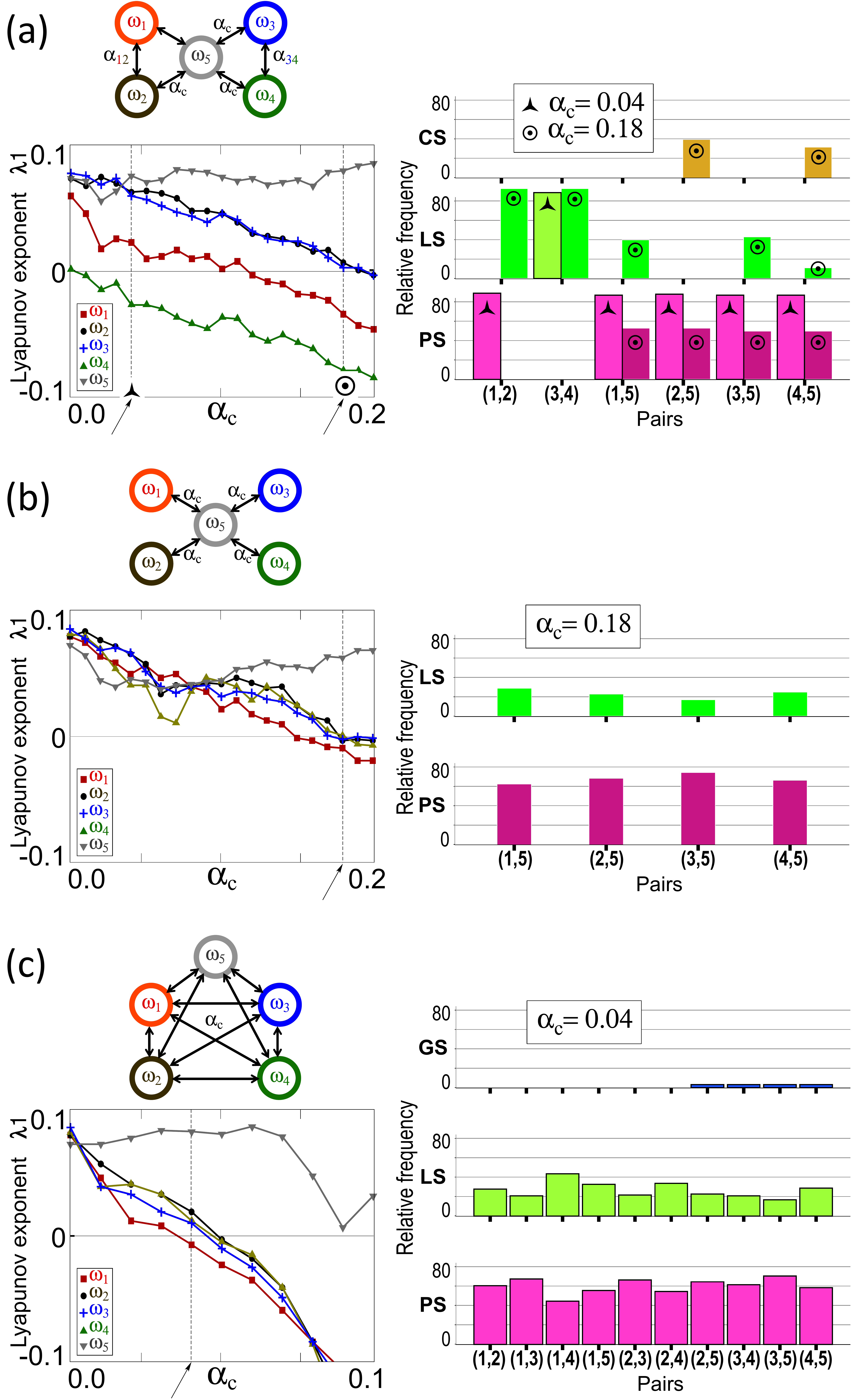}}
\caption[Consistency of the coexistence of synchronizations.]{(Color online)  
{\bf Consistency of the coexistence of synchronizations.} 
(a) Same network of Fig.{~\ref{Chap4:fig_2}(a)} with 
coupling $\alpha_{3,4}=0.20$ for node pair $3-4$.
For each node dynamics the curves show the mean value
of the maximum Lyapunov exponent ($\lambda 1$) 
as a function of coupling strength $\alpha_c$
(see Fig.{~\ref{Chap4:fig_2}(b)}).
The histogram shows the relative frequency of the synchronization patterns 
for selected values of  $\alpha_c$ 
($\alpha_c=0.04$, $\alpha_c=0.18$ indicated by the arrows).
(b) Homogeneous hub network with all couplings weighted by $\alpha_c$.
The maximum Lyapunov exponent curves for each node dynamics are similar 
and the interval of $\alpha_c$ for coexistence of synchronization patterns is small.
The histogram shows the distribution of the synchronization patterns for $\alpha_c=0.18$.
(c) All-to-all network in which all couplings are weighted by $\alpha_c$. 
The interval for coexistence of synchronization patterns is also small
and occurs for smaller values of $\alpha_c$.
The histogram shows the distribution of the synchronization patterns for $\alpha_c=0.04$.
}
\label{Chap4:fig_4}
\end{figure}

Functional networks can be constructed by establishing relationships between their (coupled)
elements. One of the most prominent dynamical features {that} functionally relates two oscillators is
synchronization, which may take the aforementioned forms (PS, GS, LS, CS) among others not studied
here. Thus, synchronization is a probe for assessing a (non) trivial relationship between
two dynamical systems. Here we want to show how the statistics of coexistence may reveal a complex
functional organization of synchronization within a network and, therefore, may help to construct
robust functional networks.

\begin{figure}[ht!]

\centering
{\includegraphics[width=\columnwidth, trim=0cm 0cm 0cm 0cm, keepaspectratio,clip=True]
{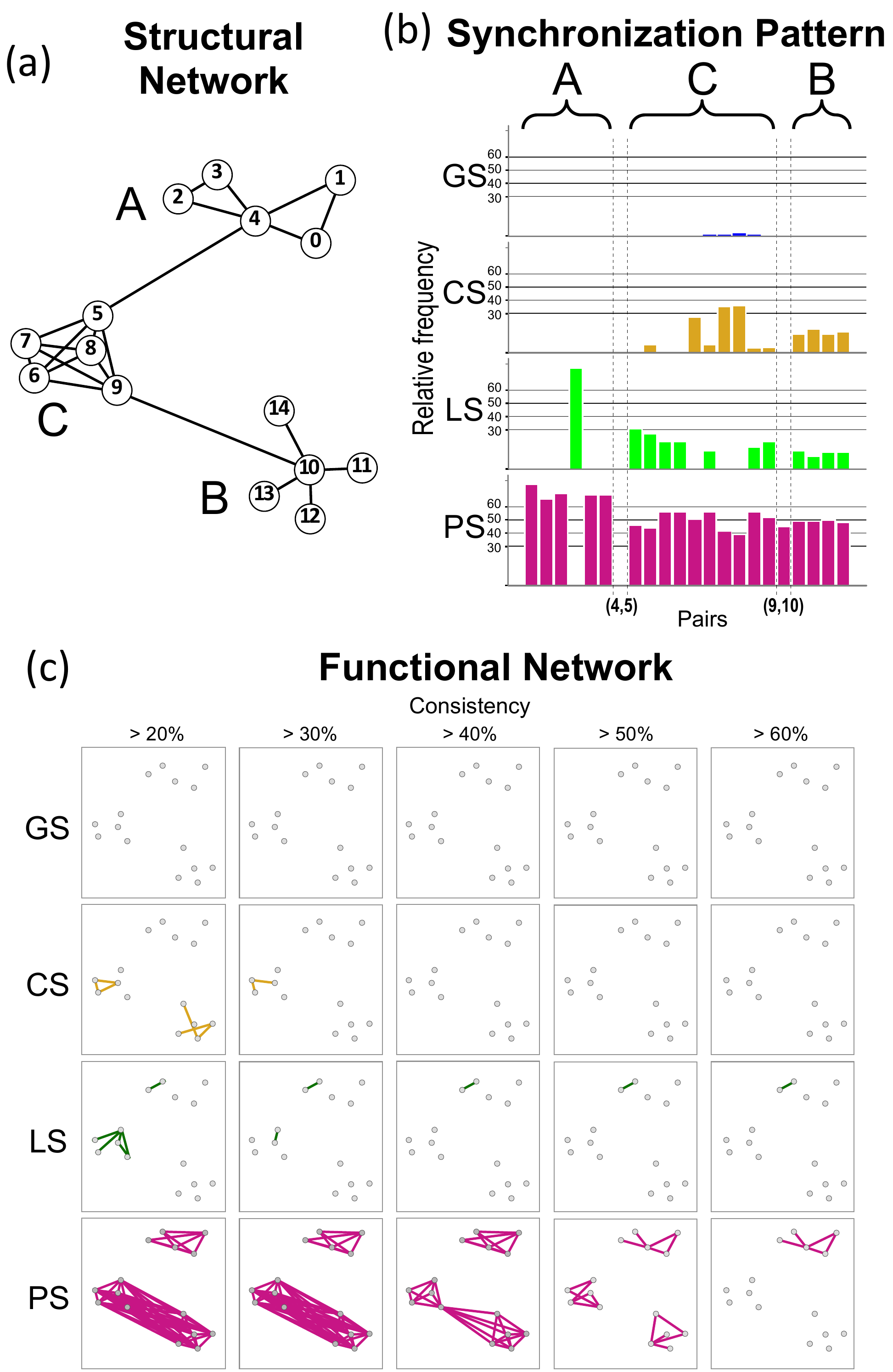}}
\caption[Construction of functional networks.]{(Color online)   
{\bf Construction of functional networks}. 
(a) A structural network of modules is constructed
by linking a module A, corresponding to a peripherally coupled network (Fig.{~\ref{Chap4:fig_4}(a)}),
with a module C, corresponding to an all-to-all network (Fig.{~\ref{Chap4:fig_4}(c)}),
by means of node pair $4-5$ with coupling $\alpha_{4,5}=0.03$
and by linking the module C
with a module B, corresponding to an homogeneous hub network (Fig.{~\ref{Chap4:fig_4}(b)}),
by means of node pair $(9,\!10)$ with coupling $\alpha_{9,10}=0.04$.
All other couplings are set equal to $\alpha_c = 0.04$.
(b)
Histogram of the relative frequency of the synchronization patterns 
for all intra-module and for the two intermodule node pairs.
Notice that the consistency of synchronization patterns in 
modules B and C
is different from the consistency observed in isolated networks
with the same topology corresponding to modules B and C (see Fig.{~\ref{Chap4:fig_4}(b),(c)}).
NS \pons{is} not shown in the histograms.
(c) Functional networks can de determined on the basis of various threshold levels
of consistency (between 20\% and 60\%)
for each type of synchronization pattern.
}
\label{Chap4:fig_5}

\end{figure}

Firstly, we take the motifs studied in Fig.{~\ref{Chap4:fig_4}} and construct a network by
connecting these groups of nodes through their hubs (or most connected nodes). The resulting graph
is shown in Fig.{~\ref{Chap4:fig_5}(a)}, where each of the motifs is labeled as A, B or C. The
intra-motif weights are the same as the selected in Fig.{~\ref{Chap4:fig_4}(a-c)},
respectively, whereas the inter-hub links weights are shown in {the caption of}
Fig.{~\ref{Chap4:fig_5}}. Figure{~\ref{Chap4:fig_5}(b)} shows the statistics of synchronization
occurrence in this network: cluster A shows a very robust consistency of its synchronizations
whereas clusters B and C are much less consistent, i.e., they display a wide repertoire of different
synchronization motifs depending on the initial conditions. However, as can be noticed when
comparing {the relative-frequency plots shown in} Figs.{~\ref{Chap4:fig_4}(a-c)} and \ref{Chap4:fig_5}(b), the dynamics of synchronization is altered when the three motifs are 
embedded in a larger network. This fact is a signature for assessing that the dynamics of coexistence in the large network is not just the simple juxtaposition of the dynamics of its composite sub-network motifs.
We now perform the task of constructing the functional networks arising from the synchronization patterns in this network. For such purpose we obviate the structural network and we make use of the statistical occurrence of each synchronization among pairs of nodes of the system. Indeed, thanks to the discrimination between each type of synchronization we can better characterize the most salient synchronization motifs between the nodes. If we establish thresholds in the statistical occurrence for each pairwise synchronization, we can extract the links which, statistically, appear the most and so are more consistent. Thus, functional networks may be constructed by taking into account the consistency of each type of synchronization among pairs of nodes.
Figure{~\ref{Chap4:fig_5}(c)} shows the construction of the functional network emerging from the structural motif-based network by applying different levels of consistency for each synchronization pattern. {For each threshold,} this construction takes into account links {that} show {\it the same} synchronization a number of times equal or larger than the consistency threshold. Accordingly, the constructed functional network coincides with the most consistent motif. This result may seem trivial as the conditions imposed in the network lead to the desired results. However, they apply to (larger) networks of coupled chaotic units.

We now take the SF prototypical network shown in Fig.~\ref{Chap4:fig_1} and perform topological changes by taking clustering as a control parameter. Figure{~\ref{Chap4:fig_6}(a)} shows the fraction of connected synchronized pairs in the SF networks whose consistency is above a certain threshold for increasing clustering. Noticeably, only low clustering networks have edges whose synchronization is consistent above a 50\% of the realizations. Therefore, only low clustering SF networks are heterogeneous enough to hold consistent synchronized dynamics. Figure{~\ref{Chap4:fig_6}(b)} shows an example of a very consistent realization-averaged SF network with clustering $C=0.15$ and a consistency map displaying the statistics of synchronization for each pair of nodes in the network. According to the statistics, the realization-averaged colors in the network mostly coincide with pure synchronization colors. Figure{~\ref{Chap4:fig_6}(c)} shows a low consistency realization-averaged SF with clustering $C=0.40$. The 
consistency map, performed for every pair of nodes in this network, shows no pattern compared to the case in panel (b). Such patterns denote that the functional organization of these networks is robust in the first case, whereas for the network with larger clustering randomized functional relationships are established among pairs of (connected) nodes.
\begin{figure}[ht!]

\centering
{{\includegraphics[width=\columnwidth, trim=0cm 0cm 0cm 0cm, keepaspectratio,clip=True]
{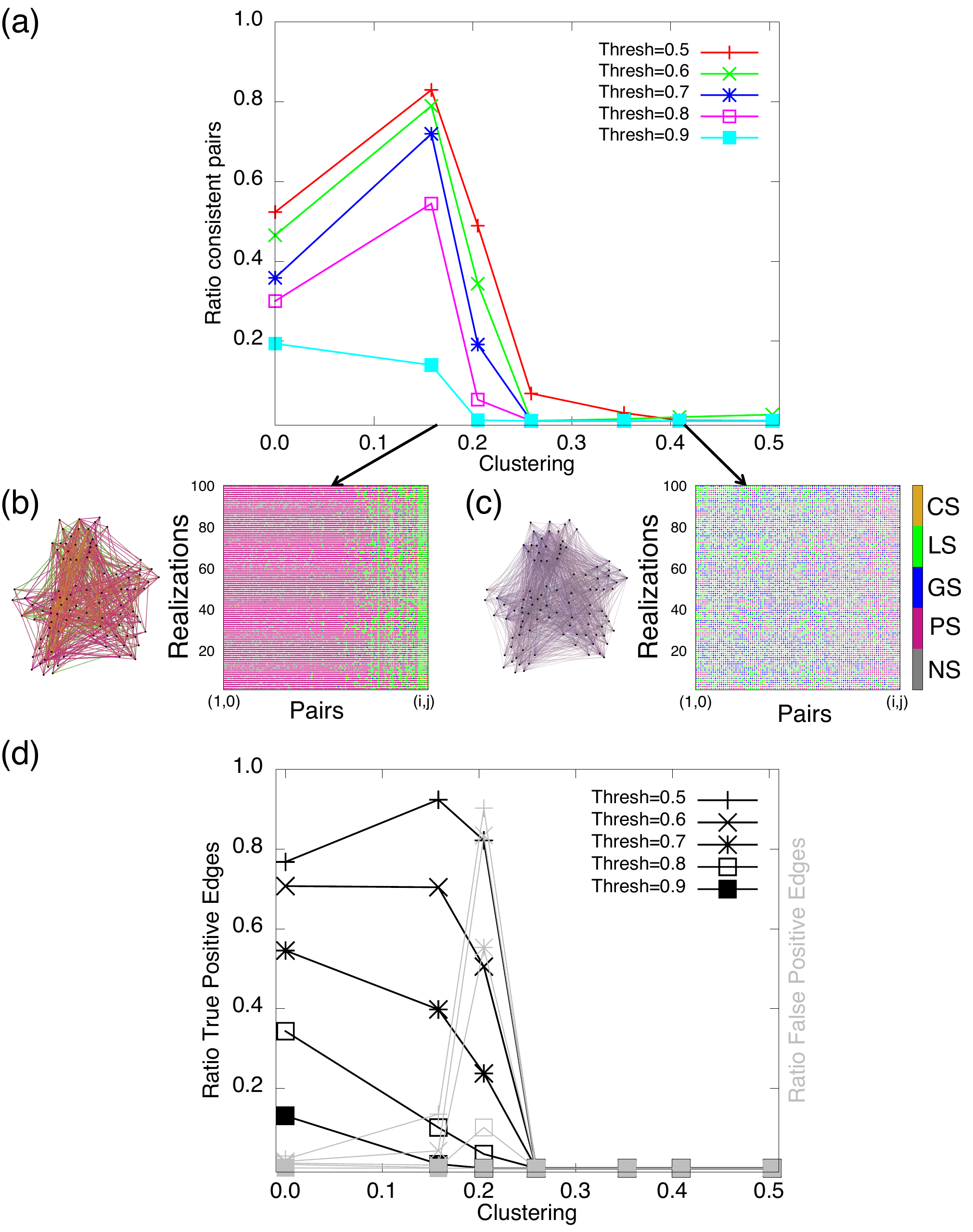}}}
\caption[Relationship between structural and functional networks for increasing clustering.]{(Color online) {\bf Relationship between structural and functional networks for increasing clustering.} (a) Ratio of consistent pairs for increasing clustering and increasing consistency thresholds ({\it Thresh}). The ratio of consistent edges displays a maximum for low clustering values, showing the dependence of this feature on the symmetries of the networks. (b) Low clustering networks ($C=0.15$) show consistent synchronization motifs, as shown in the synchronization-averaged network -- which shows almost pure synchronization colors --, and in the realization vs. pair synchronization map  -- which displays patterns of synchronization --. (c) For larger clustering networks ($C=0.40$), the synchronization-averaged networks show a single color and no patterns can be discerned in the realization vs pair synchronization map. (d) The combination of coexistence and its consistency allows to reconstruct functional 
networks that embed information of the underlying structural network. The ratio of true and false positive edges for the same networks as in panel A shows that low clustering structural networks can be reconstructed more reliably than higher clustering structural networks. In this regard, heterogeneous networks are more consistent in the synchronization dynamics and so may be easily found when extracting functional networks.}
\label{Chap4:fig_6}

\end{figure}
We analyze how many of the constructed edges are true or false positives, i.e., we quantify the matching between the functional and the structural network. We perform the following calculation:
\begin{eqnarray}
n_{true} &=& \frac{n_{c}}{n_{T}} \\
n_{false} &=& \frac{n_{in}}{n_{a-t-a} - n_{T}},
\end{eqnarray}
where $n_{T}$ is the number of edges in the structural network, $n_{a-t-a}$ is the number of edges in an equivalent all-to-all network,  $n_{c}$ is the number of constructed edges that belong to the structural network, $n_{in}$ is the number of constructed edges that do not belong to the structural network, $n_{true}$ is the ratio of constructed edges that belong to the structural network and $n_{false}$ is the ratio of constructed edges that do not belong to the structural network. In other words, $n_{true}$ computes how many of the structural edges have been reconstructed, whereas $n_{false}$ computes how many of the non-structural edges have been reconstructed.  Note that the sum $n_{true}+n_{false}$ is not equal to 1 necessarily. In this sense, a construction with high $n_{true}$ and high $n_{false}$ indicates that the constructed network is close to an all-to-all network, i.e., all structural edges can be retrieved but the number of non structural 
edges is also high, implying a bad matching between structure and function. Figure~\ref{Chap4:fig_6}(b) indicates that for clusterings below $C=0.15$ the matching between structural and functional network is high for a consistency threshold of about $50\%$, whereas the construction for higher clusterings provides either a high ratio of false positives (close to all-to-all functional network) or non-consistent networks. Interestingly, the system faces a transition point at a relatively low clustering value, $C\simeq0.21$, which prevents the construction of functional networks at higher clusterings. Indeed, as the heterogeneity in the structural network is progressively lost due to higher clustering, the system loses consistency in the synchronization motifs and so no robust functional relationships can be extracted.

The computation of true and false positives is not possible in many natural systems, e.g. the brain or signalling networks, where no exact knowledge of the anatomical structure is available. However, we raise the hypothesis that our results might unveil a potential relationship between the two if the statistics of coexistence are robust, i.e., the consistency of coexistence is high. 

\section{Conclusions}

The coexistence of synchronizations can be regarded as a phenomenon in which a variety of complex functional relationships are established between the dynamical evolutions of some coupled elements. Here we have shown that such scenario emerges in the route towards an all-synchronized network, where trivial correlations are established among oscillators. Besides, the heterogeneity in the number of node contacts and coupling strengths allows for a broad distribution of synchronization motifs. Thus, weighted networks show much more coexistence than unweighted networks.

What is more, some networks can robustly display the same coexistence patterns regardless of the initial conditions imposed{, showing high consistency}. Such feature allows to better characterize the stable functional relationships established in the network. {Besides, consistency} is at the basis of functional network construction. We argue that our method allows a better construction in terms of the statistics of synchronization motifs because we consider different coexistent synchronization states to characterize the functional network {instead of using only one}. {Finally}, we have shown that the matching between structural and functional networks is high when applying a coexistence-based reconstruction.

The consistency of {the} three prototypical networks shown in Fig.{~\ref{Chap4:fig_1}} is diverse: while SF networks with low clustering show high consistency, SW and RN networks do not display this feature because in SW or RN networks the number of node contacts fluctuates less. The consistency of the coexistence is a consequence of the heterogeneity of the network: the dynamical {synchronization} clusters consistently lay in the same {heterogeneous} synchronization manifolds for any of the initial conditions imposed {because} the synchronized trajectories are always dominated by the most connected neighbors. This allows to construct robust functional networks that have reminiscent characteristics of the structural network for increasing values of the {consistency} thresholds imposed. {Previous research shows that, in unweighted and undirected networks, {for certain} coupling regimes there is an optimal matching between structural and functional networks~\cite{Lin2015}. Here we extend these results to the case of weighted undirected networks.}
{Our results also lead us to} expect that the construction of functional networks {from real data} results in heterogeneous (non-symmetrical) networks because they are more consistent. More symmetric or homogeneous networks will appear as inconsistent if coupling is small: only when coupling is large enough to force global synchronization symmetrical networks will show up in the constructed functional networks.
On top of that, consistent dynamics eventually depends on the heterogeneity characteristics 
of the structural networks such that selected network topologies, for instance in brain dynamics,
may have been retrieved much more often than others, 
as reported elsewhere~\citep{Eguiluz2005,Bullmore2009}.

\section{Acknowledgements}
This work was partially supported 
by the Spanish Ministry of Economy and Competitiveness and FEDER (project {FIS2015-66503}). 
AEPV acknowledges support from the Swiss national Science Foundation project NEURECA (CR13I1 138032).
JGO also acknowledges support from the 
the Generalitat de Catalunya (project 2014SGR0947), the ICREA Academia programme, and from the 
``Mar\'ia de Maeztu'' Programme for Units of Excellence 
in R\&D (Spanish Ministry of Economy and Competitiveness, MDM-2014-0370).

\bibliography{PRL_Coexistence_2015.bib}

\end{document}